\documentclass[ %
 reprint,
 amsmath,amssymb,
 aps,
]{revtex4-2}
\usepackage{hyperref}
\usepackage{graphicx}
\usepackage{dcolumn}
\usepackage{bm}
\usepackage[table,xcdraw]{xcolor} 
\usepackage{graphicx} 
\usepackage{booktabs} 
\usepackage{xcolor}\definecolor{BrightOrange}{RGB}{255, 91, 0}\definecolor{DarkRed}{RGB}{139, 8, 0}\definecolor{SkyBlue}{RGB}{0, 160, 255}\definecolor{OliveGreen}{RGB}{170, 170, 0}\definecolor{BrightYellow}{RGB}{255, 198, 0}
\usepackage{transparent}

\begin{document}

\preprint{APS/123-QED}

\title{Resonance Frequency Shift Measurements of SRF Cavities at DESY}
\author{R. Ghanbari\textsuperscript{1}}
\email{rezvan.ghanbari@desy.de}
\author{T. Buettner\textsuperscript{2}}
\author{W. Hillert\textsuperscript{1}}
\author{K. Kasprzak\textsuperscript{2}}
\author{T. Krokotsch\textsuperscript{1}}
\author{R. Monroy-Villa\textsuperscript{1,3}}
\author{D. Reschke\textsuperscript{2}}
\author{L. Steder\textsuperscript{2}}
\author{A. Sulimov\textsuperscript{2}}
\author{H. Weise\textsuperscript{2}}
\author{M. Wenskat\textsuperscript{1,2}}
\author{M. Wiencek\textsuperscript{2}}
\author{J. Wolff\textsuperscript{1,2}}

\affiliation{\textsuperscript{1}University of Hamburg, Hamburg, Germany}
\affiliation{\textsuperscript{2}Deutsches Elektronen-Synchrotron DESY, Hamburg, Germany}
\affiliation{\textsuperscript{3}Johannes Gutenberg University of Mainz, Mainz, Germany}
\date{\today}

\begin{abstract}
The variation of the resonance frequency and intrinsic quality factor of superconducting \mbox{radio-frequency} cavities during the transition from the superconducting to the normal-conducting state provides essential insight into the fundamental superconducting properties of the cavity material. Investigating these transition dynamics is crucial for the continued advancement of niobium cavities whose near-surface regions are intentionally modified through the controlled introduction of interstitial atoms, such as oxygen and nitrogen, leading to the emergence of several novel behaviors whose underlying mechanisms are not yet fully understood. This work reports on the development and commissioning of a dedicated frequency-shift measurement setup. In its initial implementation, the system establishes a precise framework for determining the electron mean free path within both the superconducting penetration depth and the normal-conducting skin depth. It further enables investigation of an anomalous dip in the temperature dependence of the frequency shift near the critical temperature in cavities containing interstitial atoms in the near-surface lattice, a novel phenomenon previously reported in the literature. A recent upgrade, currently in the final stage of validation, significantly improves measurement accuracy and reproducibility. The improved setup enables comprehensive studies of the frequency shift and quality factor over the full temperature range above 7\,K, contributing to a deeper understanding of the superconducting properties.
\end{abstract}
\maketitle


\section{Introduction}
\par Niobium (Nb) superconducting radio-frequency (SRF) cavities have served as the backbone of modern particle accelerators for several decades and have demonstrated their indispensable role in accelerator technology~\cite{Reschke2017, Aune2000}. In recent years, advanced heat-treatment techniques have been developed to deliberately tailor the near-surface niobium lattice structure on nanometer-to-micrometer length scales through the controlled introduction of interstitial atoms. Among the most prominent treatments are medium-temperature (mid-T) heat treatments~\cite{Posen2020, Ito2021, Yang2023, Pan2024, Bate2024, Goedecke2025} and nitrogen doping~\cite{Bafia2019a, Gonnella2016, Fang2023}. Both approaches challenge the conventional understanding of Nb SRF cavity performance and have successfully advanced RF performance to new levels.

\par Historically, most studies of SRF cavities have focused on cryogenic temperatures below 4\,K, particularly at the operating temperature of 2\,K, by investigating the dependence of the intrinsic quality factor \(Q_0\) on the accelerating field \(E_{\mathrm{acc}}\). The intrinsic quality factor is defined as
\vspace{-5pt}
\begin{equation}
\label{Eq_01}
 Q_0 = \frac{G}{R_\mathrm{s}},
\end{equation}
where \(G\) is the geometry factor, determined by the cavity geometry and independent of material properties, and \(R_\mathrm{s}\) denotes the surface resistance. A striking and unconventional feature observed in \mbox{nitrogen-doped} and mid-T heat-treated cavities is the so-called \mbox{``\textit{anti-Q-slope}''}~\mbox{\cite{zhou2020, Posen2020, he2021}}, in which \(Q_0\) initially increases with field at moderate gradients, followed by a comparable decline at higher fields above \(\sim\!15\)~MV/m. While this effect reduces \(R_\mathrm{s}\) and consequently the RF losses, it often comes at the cost of an early quench and limits access to the highest achievable accelerating gradients~\cite{Bate2024, Steder2024, Dhakal2024, Checchin2017}. The modifications in the RF behavior of Nb cavities induced by interstitial atoms are still not fully understood and remain under active investigation, making further study essential. RF testing restricted to cryogenic temperatures below 4\,K provides only a partial picture of the superconducting behavior and is insufficient to uncover the underlying physics of niobium following these novel treatments.

\par A complementary approach is to study the behavior of relevant physical parameters over the full temperature range, from the superconducting regime through and \mbox{beyond} the superconducting-to-normal-conducting transition. Of particular importance is the temperature dependence of the penetration depth, which can be described by the two-fluid model~\cite{Gorter1934} as
\vspace{-5pt}
\begin{equation}
\label{Eq_02}
\lambda(T) = \lambda_0 \frac{1}{\sqrt{1 - \left( \frac{T}{T_\mathrm{c}} \right)^4}},
\end{equation}
where \(\lambda_0\) represents the penetration depth at zero \mbox{temperature} and \(T_\mathrm{c}\) is the critical temperature. Any variation in the penetration depth, \(\Delta \lambda\), at the inner surface perturbs the electromagnetic boundary conditions and produces a shift in the resonance frequency, \(\Delta f\). This effect can be described using the Gorter--Casimir model together with Slater’s theorem~\cite{Gorter1934, Maier1952}:
\vspace{-5pt}
\begin{equation}
\label{Eq_03}
\Delta f = \frac{-\pi \mu_0 f^2 \Delta \lambda}{G},
\end{equation}
where \(\mu_0\) is the vacuum permeability and \(f\) is the resonance frequency.

\par Another resonance-related parameter is the loaded quality factor \(Q_\mathrm{L}\), defined as
\vspace{-5pt}
\begin{equation}
\label{Eq_04}
Q_\mathrm{L} = \frac{f}{\Delta f'_\mathrm{H}},
\end{equation}
where \(\Delta f'_\mathrm{H}\) represents the full width at half maximum of the transmitted power curve. The spectral-width measurement is particularly useful for low \(Q_\mathrm{L}\) values (e.g., below \(10^8\)), a regime relevant for frequency-shift studies at temperatures above 5\,K. The loaded quality factor is related to \(Q_0\) via
\vspace{-5pt}
\begin{equation}
\label{Eq_05}
 \frac{Q_0}{Q_\mathrm{L}} = 1 + \frac{Q_0}{Q_\mathrm{ext,in}} + \frac{Q_0}{Q_\mathrm{ext,pk}},
\end{equation}
where \(Q_{\mathrm{ext,in}}\) and \(Q_{\mathrm{ext,pk}}\) are the external quality factors of the input and pickup antennas, respectively~\cite{Padamsee1998}. With \(Q_{\mathrm{ext,in}}\) and \(Q_{\mathrm{ext,pk}}\) known, the analysis of \(Q_\mathrm{L}\) enables investigation of the temperature dependence of \(R_\mathrm{s}\) above 5\,K. Therefore, monitoring the transmitted power spectra throughout the superconducting transition, from cryogenic temperatures into the normal-conducting state across \(T_\mathrm{c}\), and extracting \(\Delta f\) and \(Q_0\) provide valuable insight into the superconducting properties~\cite{Bafia2021, Herman2021}.

\par Interestingly, cavities subjected to nitrogen doping and mid-T heat treatment exhibit an additional anomalous feature: the resonance frequency as a function of temperature, \(f(T)\), shows a local minimum, the so-called ``\textit{dip},'' in the vicinity of \(T_\mathrm{c}\)~\cite{Bafia2019a, Bafia2019b, Bafia2021, Bafia2022, Ghanbari2023, Moskaitis2024, Dhakal2024, Ito2021}. This phenomenon is relatively new, and its physical origin remains unclear. The emergence of the dip further reinforces the importance of systematic frequency-shift studies, as it reveals subtle modifications of the RF response associated with the introduction of interstitial atoms into the Nb lattice structure.

\par Motivated by these objectives, we established a resonance-frequency test setup as the final step of the RF testing sequence for cavities in the vertical test cryostats at DESY~\cite{Bozhko2010, Polinski2014, Wenskat2017}. To date, this setup has been implemented and further developed for studies of mid-T heat-treated single-cell 1.3\,GHz TESLA-type Nb cavities. Mid-T heat treatment is known to dissociate the native niobium oxide layers on the niobium surface, allowing oxygen atoms to diffuse into the bulk and \mbox{introducing} interstitial oxygen within the depth range relevant for magnetic-field penetration at the cavity inner surface~\cite{Delheusy2008}. This report presents the technical aspects, challenges, and systematic progress of the frequency-shift measurements conducted by our R\&D team.

\section{Resonance-Frequency Test Setup and Methodology}
\par
\begin{figure} 
\centering
\includegraphics[scale=0.348
]{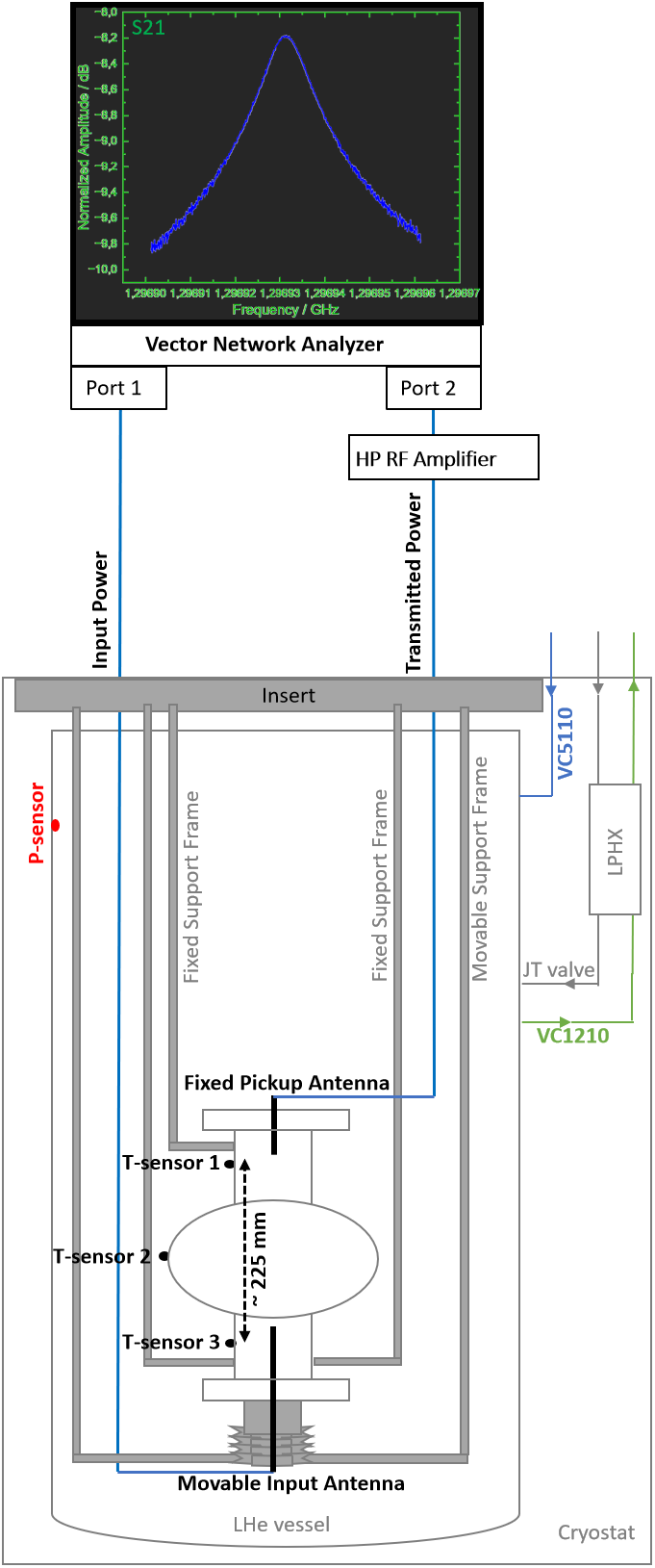}
\caption{\label{Fig_01}Experimental setup for frequency-shift measurements. Liquid helium (LHe) is removed from the vessel immediately before the measurement. During warm-up, warm helium gas (GHe) is injected through the VC5110 valve, while the VC1210 return pipe is used for gas evacuation and pressure regulation. The VC1210 is originally part of the heat-exchange system, including the low-pressure heat exchanger (LPHX) and the Joule--Thomson (JT) valve, used for cooling below the helium boiling point at atmospheric pressure (\(\sim 4.2\,\mathrm{K}\)). A pressure sensor (P-sensor) is mounted inside the cryostat, and three Cernox\textsuperscript{TM} temperature sensors (\mbox{T-sensors} 1, 2, and 3) are attached to the outer cavity surface. The pressure is kept constant during \mbox{warm-up}. \mbox{T-sensor} 2 defines the cavity temperature, while T-sensors 1 and 3 are used to evaluate the spatial temperature gradient. The input antenna is fully extended to its maximum penetration depth, corresponding to an external quality factor of approximately \(10^{8}\). A vector network analyzer (VNA) provides a 10~dBm low-power signal, and the transmitted power is amplified by a 20~dBm Hewlett--Packard (HP) RF amplifier. The S21 transmission spectra and the corresponding temperatures are recorded every minute during warm-up.}
\end{figure}

 The resonance frequency is measured during a slow, non-adiabatic warm-up from \(5\,\mathrm{K}\) to \(12\,\mathrm{K}\). This process begins with the removal of liquid helium from the cryostat, followed by the introduction of a fixed flow rate of warm helium gas. A schematic of the cryostat and the commissioned frequency-shift measurement setup is shown in Fig.~\ref{Fig_01}. To ensure that the observed frequency shift arises solely from temperature variations, rather than from deformations caused by mechanical stresses or pressure fluctuations, the pressure inside the cryostat is maintained constant throughout the measurement process~\cite{Maier1952}. This is achieved by stabilizing the pressure sensor located in the upper inner section of the cryostat through automatic regulation of the gas return line while warm helium gas is being injected. For temperature monitoring, three sensors are installed on the outer surface of the cavity---at the top, equator, and bottom---to continuously record the temperature during the experiment. The equatorial temperature is used to analyze the temperature dependence of the resonance frequency and quality factor, while the readings from the top and bottom sensors are used to evaluate spatial temperature gradients.

\par
\begin{figure*}
\includegraphics[scale=0.58]{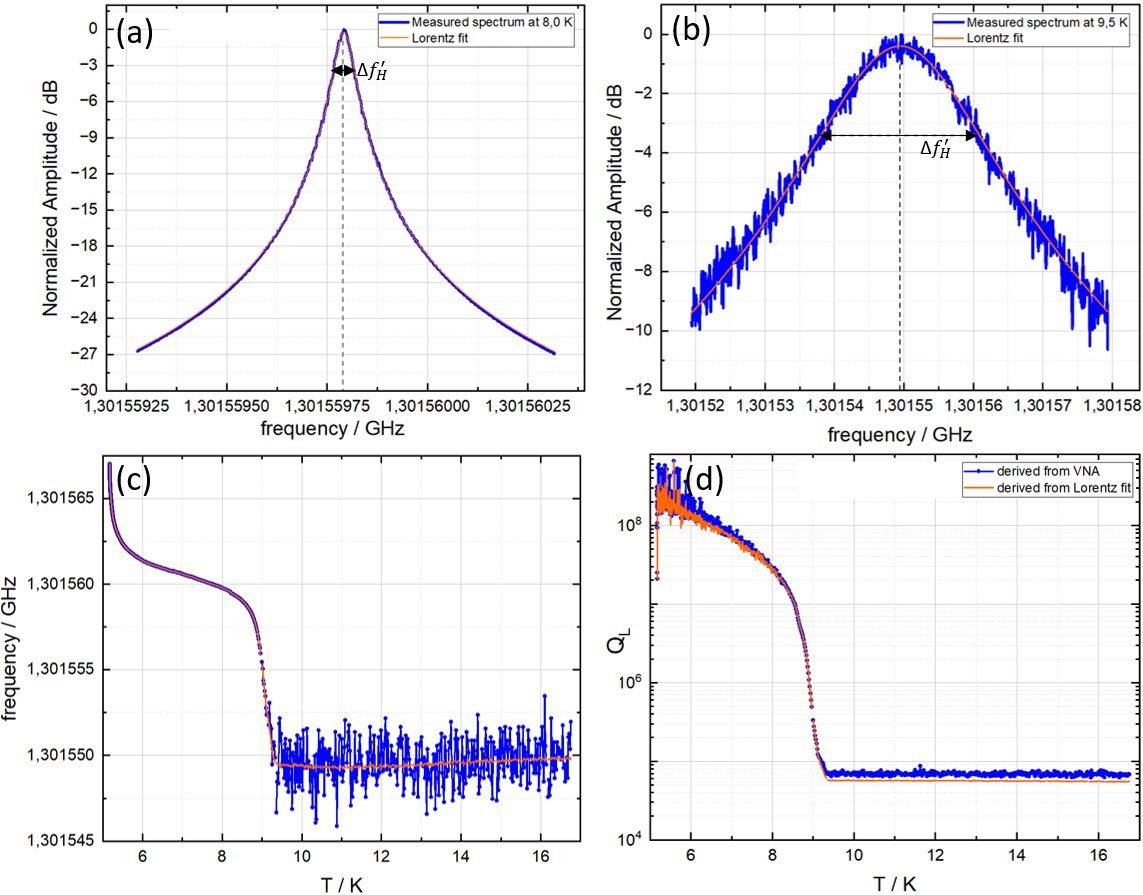}
\caption{\label{Fig_02}Accuracy enhancement of resonance-frequency analysis through Lorentzian fitting for an exemplary cavity. 
Example spectra: (a) below the critical temperature, exhibiting a low noise floor, and (b) above the critical temperature, showing an increased noise level. This elevation markedly increases the uncertainty in determining both the resonance peak frequency and the \(-3\)\,dB bandwidth \(\Delta f'_\mathrm{H}\) used for calculating \(Q_\mathrm{L}\).
Applying a Lorentzian fit effectively mitigates the adverse effects of noise on both peak identification and bandwidth estimation. The extracted resonance frequencies and the corresponding loaded quality factors \(Q_\mathrm{L}\) are shown in (c) and (d), respectively, obtained from vector network analyzer (VNA) measurements and subsequent Lorentzian fitting as a function of cavity temperature \(T\).
The point-wise statistical error of the VNA (IFBW = 10\,Hz, 601 points, 10\,dBm) is considered negligible compared to the noise floor. 
The statistical error of the Cernox\textsuperscript{TM} temperature sensors is on the order of a few millikelvin and is likewise considered negligible. 
With the application of Lorentzian fitting and minimization of noise-floor effects, the uncertainties in the extracted resonance frequency and \(Q_\mathrm{L}\) values in the normal-conducting state are reduced to approximately 40\,Hz and 0.003, respectively, and can therefore be considered negligible. 
At lower temperatures, the uncertainty in the \(\Delta f'_\mathrm{H}\) determination increases, and its influence on \(Q_\mathrm{L}\) is observed as fluctuations.}
\end{figure*}

\begin{figure*}
\includegraphics[scale=0.41]{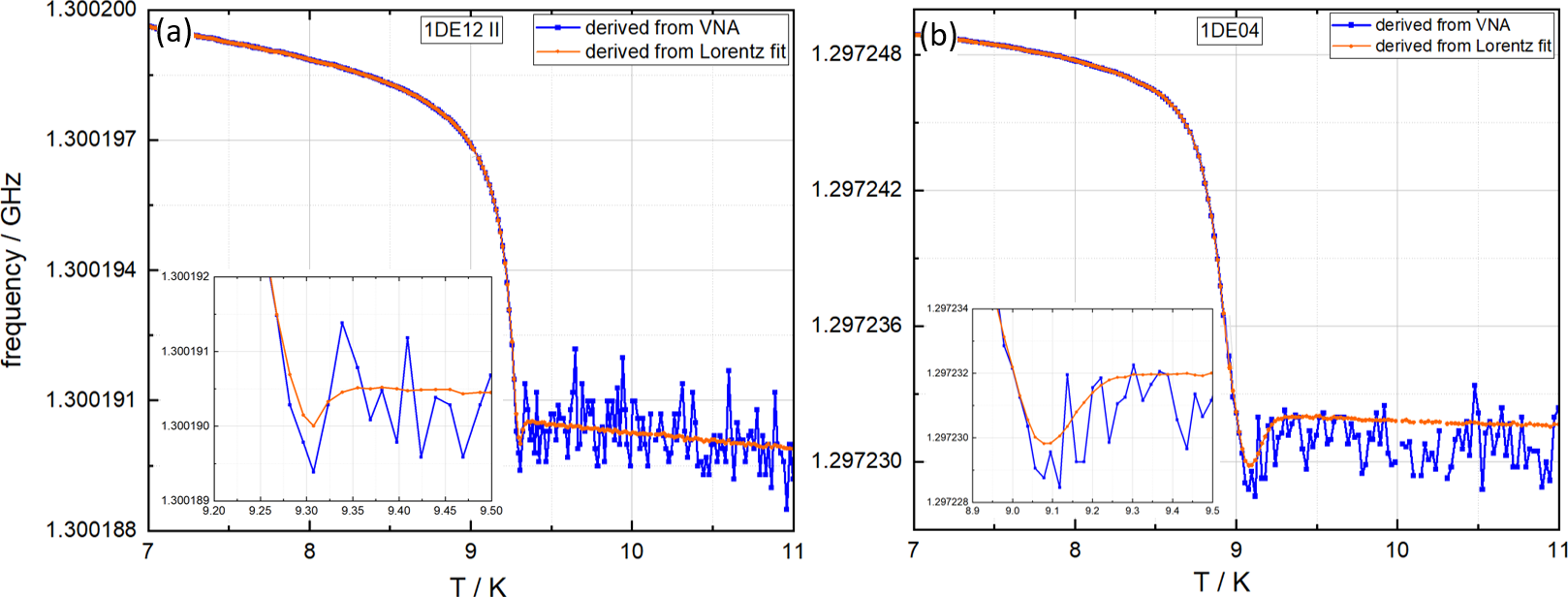}
\caption{\label{Fig_03}Two exemplary cavities, shown in (a) and (b), illustrate the importance of Lorentzian fitting for identifying and accurately measuring the resonance-frequency dip around the critical temperature in the cryogenic temperature \(T\) dependence of the resonance frequency for mid-T heat-treated cavities. Around and above the critical temperature, elevated noise levels obscure the accurate determination of the resonance frequency directly from the vector network analyzer (VNA). The insets show a zoom-in of the dip around the critical temperature.}
\end{figure*}

 To measure the resonance frequency and loaded quality factor of the cavities, transmitted power spectra are recorded every minute using a vector network analyzer (VNA). As shown in Figs.~\ref{Fig_02}(a) and (b), the resonance frequency and the full bandwidth are obtained by fitting each spectrum with a Lorentzian function,
\begin{equation}
\label{Eq_06}
P(f') = |A(f')|^2 = A_0^2 \cdot \frac{(\Delta f'_\mathrm{H}/2)^2}{(f' - f)^2 + (\Delta f'_\mathrm{H}/2)^2},
\end{equation}
where \(P(f')\) denotes the signal power as a function of frequency \(f'\), and \(A_0\) represents the peak amplitude. This choice is well justified, since an SRF cavity can be modeled as a driven, damped harmonic oscillator whose squared-amplitude response follows a Lorentzian distribution near resonance. The fitting procedure is \mbox{particularly} effective above \(T_\mathrm{c}\), where the spectra exhibit \mbox{elevated} noise levels, as shown in Fig.~\ref{Fig_02}(b). The resulting improvements in the extracted \(f(T)\) and \(Q_\mathrm{L}(T)\) \mbox{values} obtained through Lorentzian fitting are illustrated in Figs.~\ref{Fig_02}(c) and (d). The significance of the Lorentzian approach, further demonstrated in Fig.~\ref{Fig_03}, lies in its ability to robustly identify and characterize the \mbox{frequency-dip} features observed in the vicinity of \(T_\mathrm{c}\) in mid-T \mbox{heat-treated} cavities.

\begin{figure*}
\centering
\includegraphics[scale=0.372]{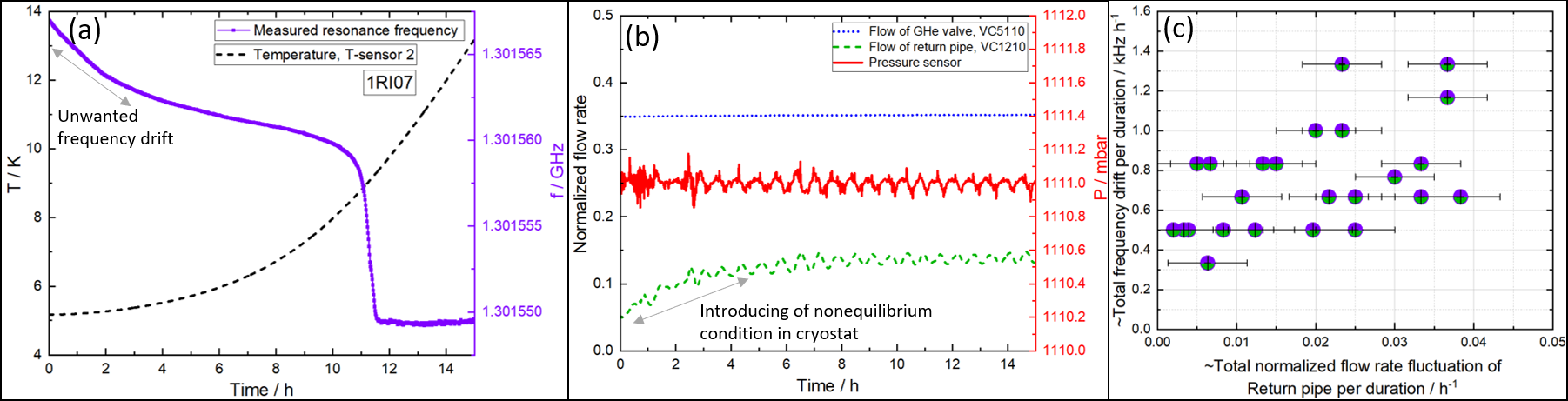}
\caption{\label{Fig_04}Monitored parameters during the cryostat warm-up for resonance-frequency-shift measurements:  
(a) Resonance frequency \(f\) and cryogenic temperature \(T\) of the cavity as a function of time during warm-up. A frequency drift at the onset of the measurements is observed, which, based on theory, is not due to the temperature dependence of the penetration depth and, consequently, the resonance-frequency shift. The unwanted frequency drift typically starts at the lowest measured temperature and gradually decreases at higher temperatures.  
(b) Corresponding pressure \(P\) read from the P-sensor, maintained at \(1111 \pm 0.16\,\mathrm{mbar}\), and flow rates, each normalized to its respective maximum flow capacity, as a function of time. The VC5110 valve is used for warm helium gas (GHe) injection, and the VC1210 return pipe is used for gas evacuation and pressure control. The frequency drift is hypothesized to result from an initial mechanical force exerted on the cavity by the insert due to thermal expansion of the insert under non-equilibrium conditions within the cryostat at the onset of the warm-up process, while the return pipe is being regulated until the valves stabilize, maintaining constant pressure under thermal equilibrium in the cryostat.  
(c) The averaged initial change in evacuation flow through the return pipe (VC1210) per unit time and the corresponding averaged initial resonance-frequency drift per unit time in all measurements demonstrate an indirect correlation between the unwanted frequency drift and cavity expansion arising from thermal non-equilibrium conditions.}
\end{figure*}

\begin{figure*}
\includegraphics[scale=0.53]{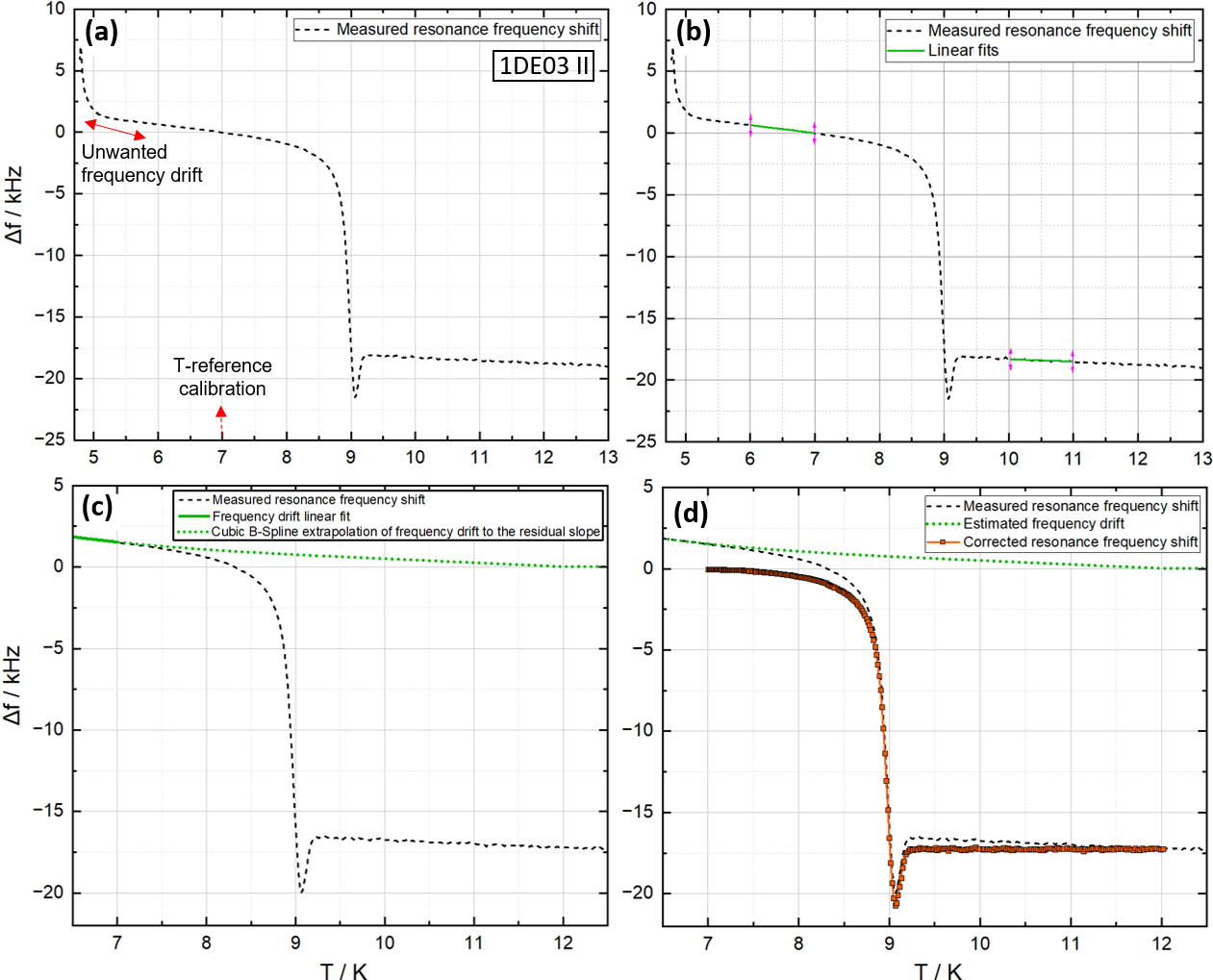}
\caption{\label{Fig_05}Systematic empirical frequency-drift correction:  
(a) An unwanted frequency drift is observed at the onset of the frequency-shift measurement.  
(b) The frequency-drift slope is determined at the start of the measurement and at the residual drift level above critical temperature \(T_{\mathrm{c}}\). (c) The first linear fit at \(7\,\mathrm{K}\) is extrapolated to match the slope of the second linear fit above \(T_{\mathrm{c}}\), in order to empirically estimate the frequency-drift curve.  
(d) The extrapolated frequency-drift curve is subtracted from the measured resonance frequency to obtain the corrected resonance-frequency shift. Consequently, only the corrected resonance-frequency shifts acquired above \(7\,\mathrm{K}\) were included in the final analysis, corresponding to the temperature range theoretically sufficient for the intended evaluation.}
\end{figure*}

\begin{figure*}
\includegraphics[scale=0.59]{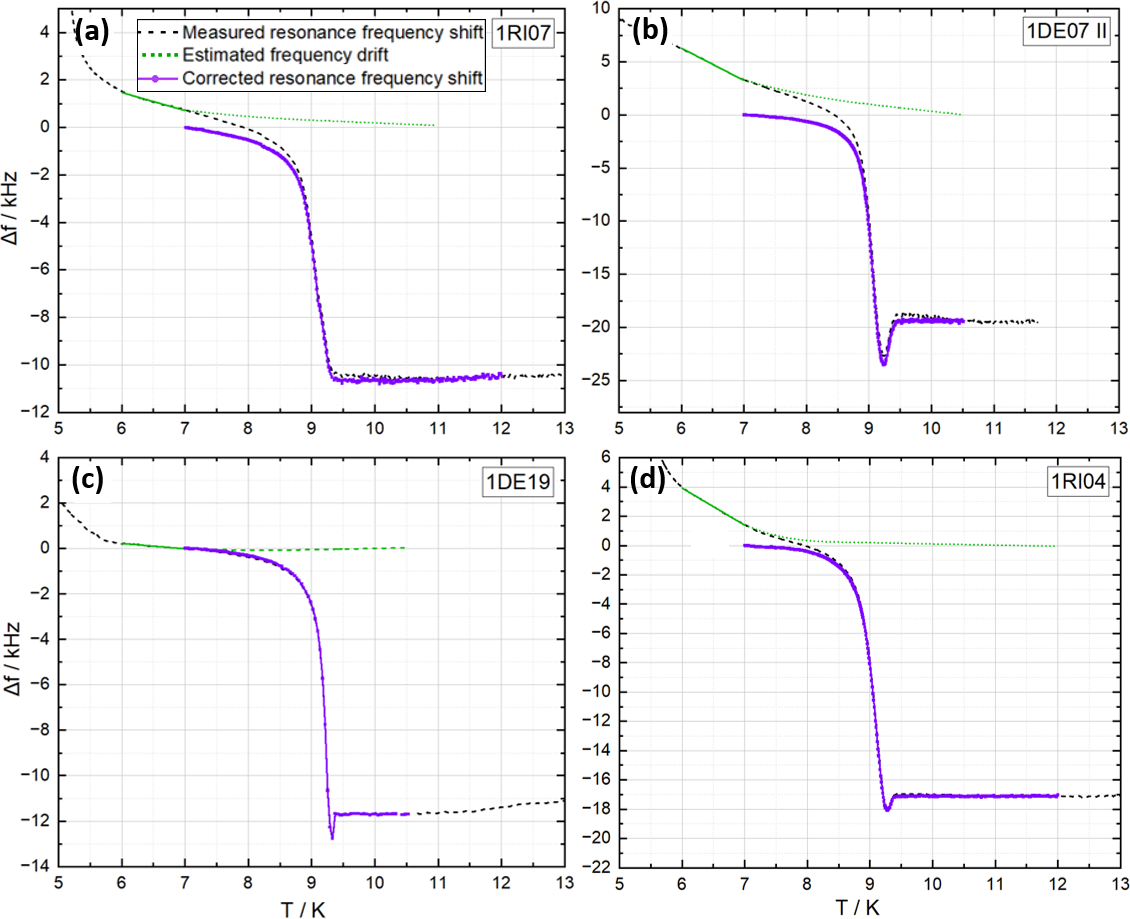}
\caption{\label{Fig_06}
Representative examples of post-measurement empirical frequency-drift corrections applied systematically to the data. Based on repeated measurements, the maximum relative uncertainties in the corrected frequency shifts (purple solid curves with data points and error bars) are estimated to be approximately \(25\%\) of the empirical frequency drifts (green dotted curves). A systematic temperature offset of approximately \(0.11\,\mathrm{K}\) in the temperature range around \(T_\mathrm{c}\) was identified between two T-sensor~2 positions at the cavity equators in two different inserts used for the measurements. This offset was corrected, and a corresponding systematic error of \(0.11\,\mathrm{K}\) was included in the uncertainty analysis. For measurements conducted after a specific date, when the T-sensor positioning accuracy was within \(\pm 6\)\,mm, the maximum relative cavity-temperature uncertainty due to spatial temperature gradients was estimated to be approximately \(10\%\) of the measured temperature difference between the top and bottom of the cavity, as recorded by T-sensor~1 and T-sensor~3 in Fig.~\ref{Fig_01}. A higher uncertainty of up to \(14\%\), observed in some earlier measurements, resulted from imprecise positioning of T-sensors~1 and~3, with a separation of 225\,mm (\(-62/+12\)\,mm). All results presented in this paper account for the corresponding maximum relative uncertainties in both the frequency shift and the cavity temperature.}
\end{figure*}

\par The resonance frequency and monitored parameters related to the slow warm-up during the frequency-shift measurement are presented in Fig.~\ref{Fig_04}. In part (a), a representative example of resonance-frequency changes alongside the increase in cavity temperature is shown. Part (b) displays the cryostat conditions, including pressure, a fixed warm-helium-gas injection flow rate, and the helium-gas evacuation flow rate during a typical \mbox{warm-up} process for the frequency-shift measurement. During the measurements, the pressure sensor maintained a \mbox{stable} pressure of \(1111\,\mathrm{mbar}\), exhibiting minimal fluctuations with a root-mean-square error of \mbox{\(\sim 0.16\,\mathrm{mbar}\)}. The \mbox{impact} of these pressure fluctuations on the resonance frequency is negligible, and the measured \mbox{Lorentzian-fitted} \(f\), as shown in Fig.~\ref{Fig_04}(a), remains free of any noticeable \mbox{root-mean-square-level} fluctuations.

\par The temperature dependence of the penetration depth follows a fourth-power law, as described in Eq.~\ref{Eq_02}, which, according to Eq.~\ref{Eq_03}, enhances the sensitivity of the resonance frequency to increasing temperature and leads to a pronounced shift as the temperature approaches \(T_{\mathrm{c}}\). In theory, approximately \(95\%\) of the total change in penetration depth---and consequently the shift in resonance frequency---occurs between \(7\,\mathrm{K}\) and \(T_{\mathrm{c}}\). Below \(7\,\mathrm{K}\), the resonance frequency becomes effectively temperature-independent, and below \(5\%\) of the frequency shift is expected to occur in this lower temperature range~\cite{Gorter1934}. However, undesired, gradually diminishing frequency drifts were detected at the onset of the \(f(T)\) measurements below \(7\,\mathrm{K}\), as shown in Fig.~\ref{Fig_04}(a). Our observations indicate that, upon initiating the injection of warm helium gas, the automated gas-return system continuously adjusts to maintain a constant pressure reading from the pressure sensor, while the gas flow rate in the return pipe gradually increases and eventually stabilizes over time during the measurement, as shown in Fig.~\ref{Fig_04}(b). Figure~\ref{Fig_04}(c) illustrates that the flow rate over which the gas-return system adjusted toward stabilization correlates with the rate of gradually diminishing frequency drifts observed at the beginning of the measurements.

\par It is hypothesized that the introduction of warm \mbox{helium} gas into the cryostat initially creates transient thermal non-equilibrium conditions within the cryostat volume. A temperature imbalance arises between the regions where warm helium gas is injected and where helium gas is evacuated through the return pipe, leading to thermal gradients along the insert frames. This condition leads to thermal expansion of the frames; consequently, mechanical forces from the fixed support frames between the top and bottom of the cavity (see Fig.~\ref{Fig_01}) act on the cavity volume, causing an unwanted shift in the resonance frequency of the mounted cavity until the return pipe stabilizes. Hence, thermal equilibrium throughout the cryostat volume is achieved by maintaining \mbox{steady-state} flow rates for both warm helium gas injection and helium gas return, while the pressure sensor ensures constant pressure with low \mbox{root-mean-square-level} fluctuations. To minimize this effect, the frequency-shift measurements were initiated at \(5\,\mathrm{K}\) to allow the initially observed frequency drifts to diminish gradually, enabling the system to stabilize above \(7\,\mathrm{K}\) and reach minimal residual levels near \(7\,\mathrm{K}\), which satisfies the theoretically required temperature range for the frequency-shift study.

\par To enhance analytical precision and quantify the associated uncertainty arising from the non-equilibrium conditions at the onset of the measurements---manifested as a drift in the resonance frequency shift---a systematic empirical frequency-drift correction was applied to the measured frequency shifts for all cavities after the measurements. This correction involved subtracting an estimated drift, determined empirically for each \(\Delta f(T)\) curve, as outlined step by step in Fig.~\ref{Fig_05}. To further illustrate the impact of the frequency-drift correction, representative examples of the corrected \(f(T)\) curves for different \mbox{mid-T} \mbox{heat-treated} cavities are presented in Fig.~\ref{Fig_06}. Since the frequency drift varies from one measurement to another, the measurements are not directly reproducible. Applying the frequency-drift correction to identify and remove the unwanted drift in the temperature dependence of the resonance-frequency shift enhances the comparability of the resonance-frequency measurements. Another source of measurement non-reproducibility arises from the spatial temperature gradient along the cavity, which varies between measurements and affects the cavity temperature distribution from iris to iris, thereby influencing the accuracy of the cavity-temperature reading determined at the equator.

\par To account for uncertainties arising from both mechanical stresses acting on the resonance frequency as frequency drifts and spatial temperature gradients along the cavity, three repeated frequency-shift measurements were performed for each of two cavities before mid-T heat treatment and for one cavity after mid-T heat treatment. Our observations indicate that the maximum relative uncertainties of the corrected frequency shifts are approximately \(25\%\) of the empirically determined \mbox{frequency} drifts. The maximum relative uncertainties of the \mbox{cavity-temperature} readings are estimated to be approximately \(10\%\)---and up to \(14\%\) for some cavities with uncertain T-sensor positioning---of the measured temperature differences between the top and bottom of the \mbox{cavities} (T-sensor~1 and T-sensor~3 in Fig.~\ref{Fig_01}). Both \mbox{uncertainties} are included in Fig.~\ref{Fig_06} and in all subsequent figures as error bars throughout this report.

\par Achieving more reliable frequency-shift measurements by reducing uncertainty and addressing its sources remains an ongoing objective of our team. Two recent improvements in the measurement procedure, which have resulted in a significant reduction in uncertainty, are discussed in the following section.
\section{Recent Measurement Improvements}
\begin{figure}
\includegraphics[scale=0.36]{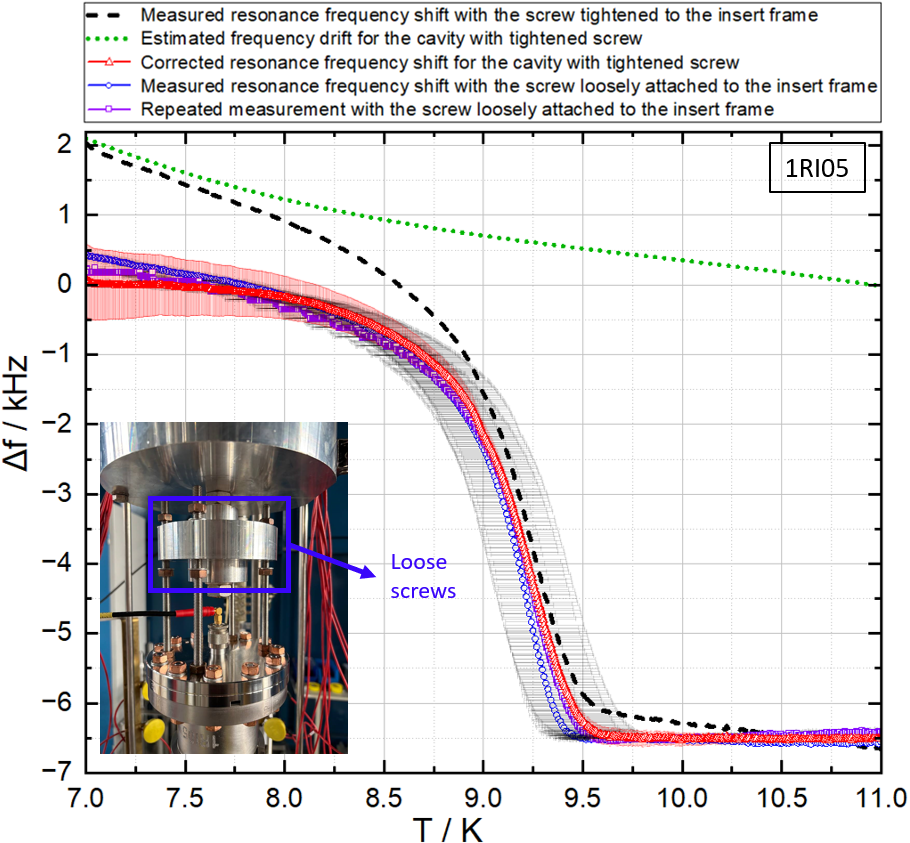}
\caption{\label{Fig_07}Three repeated frequency-shift measurements of the same cavity. One test was performed with the mounting screw connecting the cavity to the insert fully tightened (dashed curve), resulting in a frequency artifact. Two subsequent tests were conducted with the screws intentionally left slightly loose, as shown in the inset. This additional mechanical freedom relieved stress and reduced cavity deformation caused by insert thermal expansion under transient non-equilibrium conditions (see Fig.~\ref{Fig_04}). Consequently, the frequency artifact disappeared, and the measurements with the screw loosely attached to the insert exhibited reliable reproducibility, with an uncertainty of approximately 0.5\,kHz at 7\,K. The \mbox{frequency-shift} curve obtained with the tightened screw was corrected using the systematic empirical drift-correction method developed in this work (see Fig.~\ref{Fig_05}). The corrected curve, including its uncertainty margin, agrees closely with the loose-screw measurements; both lie within its uncertainty bound, confirming the validity of the empirical correction method and its uncertainty definition developed in this work. Except for this improved loose-screw configuration, all other measurements reported in this paper were performed with the screw fully tightened, followed by post-measurement frequency corrections. Minor discrepancies \mbox{between} the curves near \(T_\mathrm{c}\) are attributed to different spatial temperature gradients during individual measurements; the apparent shift of the cavity critical temperature toward higher values corresponds to larger spatial temperature gradients.}
\end{figure}

\par A validation test was recently conducted to examine the hypothesis that thermal expansion of the insert exerts mechanical stress on the mounted cavity, thereby altering its resonance frequency at the onset of the measurements. To test this, the cavity---originally rigidly mounted to the insert---was partially released by loosening the upper mounting screws, thereby allowing limited mechanical freedom (see the inset of Fig.~\ref{Fig_07}). The \(f(T)\) results of two repeated tests with loosened screws, shown in Fig.~\ref{Fig_07}, demonstrate that the frequency drift \mbox{observed} at the beginning of the measurement nearly vanished after the screws were loosened. This observation supports the proposed origin of the frequency drift observed in the measured frequency shifts. Furthermore, this \mbox{approach} offers a practical method for significantly improving frequency-shift measurements, yielding stable, reproducible, and reliable data with a substantial reduction in systematic error for future experiments.

\par This test also provided an opportunity to verify the reliability of the empirical frequency-drift correction method applied to previously measured data. As shown in Fig.~\ref{Fig_07}, when the same cavity was measured with the screws tightened, the onset frequency drift reappeared. Applying the systematic empirical correction defined and described in this study yields a corrected \(\Delta f(T)\) curve that agrees closely with the newly obtained data from the loosened configuration. The overlap between the repeated loose-screw curves and the corrected fully tightened-screw curve, within its uncertainty margin, strongly supports the reliability of both the empirical correction and the associated uncertainty estimation applied in the \(\Delta f(T)\) analysis. This observation confirms the effectiveness of the empirical correction approach and the consistency of the previously analyzed data when the associated uncertainties are taken into account.

\begin{figure}
\includegraphics[scale=0.49]{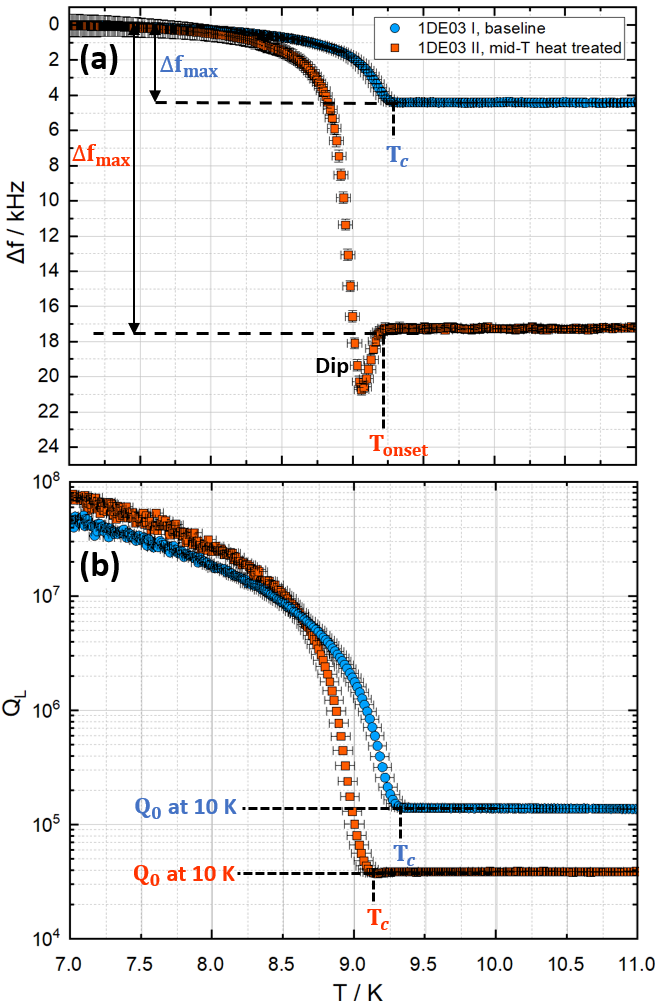}
\caption{\label{Fig_08}Cryogenic temperature dependence of (a) \mbox{resonance-frequency} shift \(\Delta f\) and (b) loaded quality factor \(Q_\mathrm{L}\) for an exemplary cavity before (\mbox{1DE03 I}) and after (\mbox{1DE03 II}) a mid-T heat treatment of 3 hours at \(300^{\circ}\mathrm{C}\). The total frequency shift \(\Delta f_{\mathrm{max}}\), critical temperature \(T_\mathrm{c}\), dip-onset temperature \(T_\mathrm{onset}\), and \(Q_\mathrm{L}\) at \(10\,\mathrm{K}\) are indicated and used in the analysis presented in this study. Due to the emergence of the dip around \(T_\mathrm{c}\), a quantitative definition of \(T_\mathrm{c}\) in the frequency shift of mid-T heat-treated cavities requires further investigation. The differences in the curve features include a larger \(\Delta f\), a suppression of \(T_\mathrm{c}\), a dip around \(T_\mathrm{c}\), a lower \(Q_\mathrm{L}\) above \(T_\mathrm{c}\), and a higher \(Q_\mathrm{L}\) at temperatures below \(8.5\,\mathrm{K}\) after mid-T heat treatment, consistent with observations reported by other laboratories~\cite{Bafia2019a, Bafia2019b, Bafia2021, Bafia2022, Moskaitis2024, Dhakal2024}.}
\end{figure}

\par The influence of the spatial temperature gradient along the cavity on the cavity-temperature reading, with the corresponding uncertainties indicated by the error bars in Fig.~\ref{Fig_07}, is evident. Larger \(\nabla T\) values lead to an apparent shift of \(T_\mathrm{c}\) toward higher values. This effect \mbox{occurs} when the temperature sensor located at the cavity equator---used to represent the cavity temperature---crosses the true critical temperature of niobium while the lower part of the cavity, from the equator to the bottom iris, remains in the superconducting state and thus continues to affect the frequency shift and quality factor. Consequently, the effect of the spatial temperature gradient on \(f(T)\) and \(Q_0(T)\) at temperatures around \(T_\mathrm{c}\) acts in a single direction, and in this temperature region the physically meaningful temperature uncertainty has only negative values, resulting in an apparent increase in the observed critical temperature. In contrast, at lower temperatures in the \(f(T)\) and \(Q_0(T)\) curves, where the entire cavity is in the superconducting state, the influence of spatial temperature gradients is reduced and may even reverse direction, producing positive uncertainty values as well. Hence, although a more accurate and physically meaningful uncertainty analysis is in principle feasible, it is practically inefficient. Therefore, minimizing the temperature gradient along the cavity during \(f(T)\) and \(Q_0(T)\) measurements remains the most effective and pragmatic approach.

\par During the frequency-shift measurements on one of the cavities, both before and after the mid-T heat treatment, the cavity was installed in an insert equipped with an additional setup. This setup was completely independent of our measurements and remained inactive during the frequency-shift tests; only its physical presence around the cavity is relevant here. In this configuration, the cavity was surrounded by electronic boards designed for B-mapping studies~\cite{Wolff2023, Bate2024}. For these frequency-shift measurements, the spatial temperature gradients were significantly reduced to a reproducible level of approximately \(2\,\mathrm{K/m}\) over the course of six well-controlled tests. \mbox{Figure~\ref{Fig_08}} shows the \(f(T)\) and \(Q_\mathrm{L}(T)\) results for this cavity (1DE03), measured before and after mid-T heat treatment, where the incidental surrounding coverage provided a more homogeneous thermal distribution along the cavity. The spatial temperature gradient and the corresponding uncertainty in the cavity temperature were maintained within a controlled and minimized range in Fig.~\ref{Fig_08}, compared to other experiments (e.g., the temperature uncertainty in Fig.~\ref{Fig_07}). This resulted in one of the most stable and reproducible measurements achieved in this study, demonstrating a high level of reliability. The corresponding uncertainties in Fig.~\ref{Fig_08} lie within a range sufficiently accurate for \(\lambda(T)\) and \(R_\mathrm{s}(T)\) analyses over the entire temperature range above 7\,K and for a precise study of \(T_\mathrm{c}\), marking a major step toward the successful completion of future frequency-shift measurements. The implementation of such an upgraded frequency-shift measurement setup, incorporating thermal coverage around the cavity during the measurement, is currently under discussion and evaluation within our group.

\par The final important consideration for achieving successful measurements concerns the effect of antenna \mbox{coupling}. With \(Q_{\mathrm{ext,pk}}\) on the order of \(10^{11}\), the fixed \mbox{pickup-coupling} term in Eq.~\ref{Eq_05} can be considered negligible for all measurements. In previous \mbox{experiments}, the maximum penetration depth of the input antenna used for the frequency-shift measurements corresponded to \(Q_{\mathrm{ext,in}}\) values on the order of \(10^8\); however, individual measurements of \(Q_{\mathrm{ext,in}}\) for each mounted \mbox{input} \mbox{antenna} were not performed or recorded with high \mbox{accuracy}. For future investigations aiming to conduct reliable \mbox{temperature-dependent} frequency and \mbox{quality-factor} studies from 7\,K to above \(T_\mathrm{c}\), the first step should be to determine \(Q_{\mathrm{ext,in}}\) at the maximum penetration depth of the input antenna prior to each \mbox{frequency-shift} experiment. This procedure would also help minimize systematic errors associated with \mbox{motor} positioning of the input antenna at its maximum \mbox{insertion} depth for each measurement. Implementing this approach will allow the complete \(Q_\mathrm{L}(T)\) curve to be converted into the corresponding \(Q_0(T)\) curve and, consequently, into a reliable \(R_\mathrm{s}(T)\).

\par As numerous measurements on mid-T heat-treated cavities had already been conducted prior to the final methodological refinements, this systematic report is structured to provide a consistent framework for incorporating and analyzing the previously acquired \mbox{resonance-frequency} data, as presented in the Appendix. It focuses on extracting physically meaningful quantities from the measured total frequency shifts below \(T_\mathrm{c}\) and quality factors above \(T_\mathrm{c}\), and on identifying reliable parameters, such as the electron mean free path and oxygen concentration, which will be used in future investigations.

\par
\begin{figure}
\includegraphics[scale=0.34]{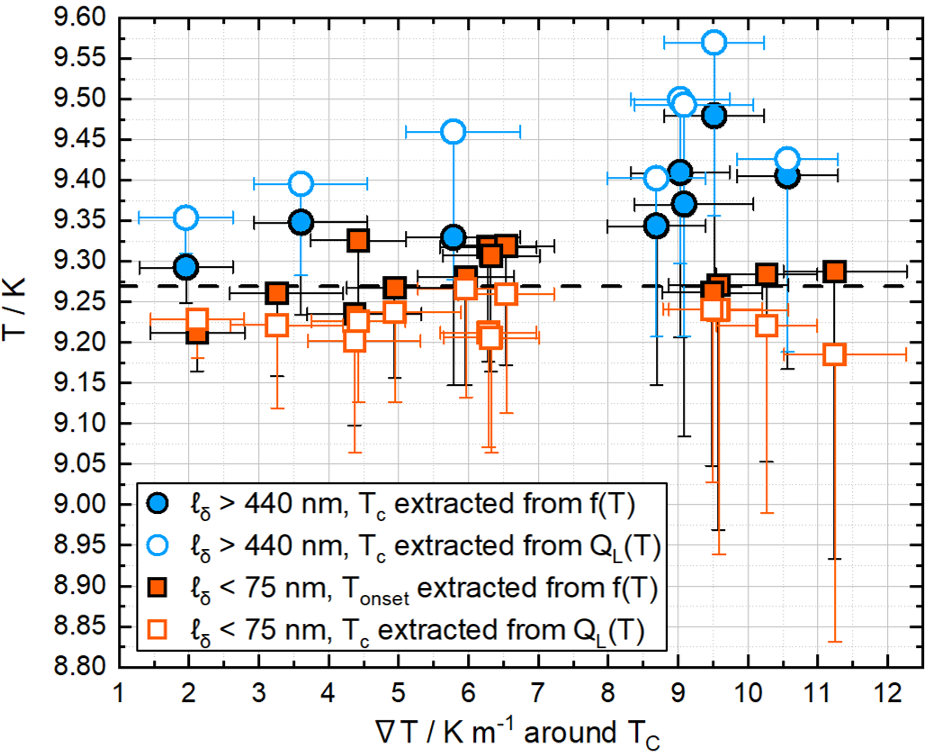}
\caption{\label{Fig_09}Effect of spatial temperature gradients \(\nabla T\) on the critical temperature \(T_\mathrm{c}\) and the dip-onset temperature \(T_\mathrm{onset}\) in \(f(T)\) and \(Q_\mathrm{L}(T)\) curves (see Fig.~\ref{Fig_08} for the definitions of \(T_\mathrm{c}\) and \(T_\mathrm{onset}\)). Such gradients hinder accurate determination of the true critical temperature, shifting it unidirectionally toward higher apparent values; the corresponding maximum uncertainties are represented as error bars. The dashed line indicates the critical temperature of niobium, \(9.27\,\mathrm{K}\). Our \mbox{observations} (currently under further investigation) show that covering around the cavity during frequency-shift measurements can mitigate direct exposure to the cryostat environment, thereby minimizing temperature inhomogeneity along the cavity and lowering the spatial temperature gradient to a reproducible and reliable range of approximately \(2\,\mathrm{K/m}\). The low-uncertainty curves in Fig.~\ref{Fig_08} illustrate exemplary successful measurements of both \(f(T)\) and \(Q_\mathrm{L}(T)\) curves over the full temperature range above \(7\,\mathrm{K}\), obtained under low spatial temperature gradients of \(\sim 2\,\mathrm{K/m}\). The weaker influence of temperature gradients on shifting \(T_\mathrm{c}\) to higher values in cavities after mid-T heat treatment (orange square symbols, \(\ell_\delta < 75\,\mathrm{nm}\)) compared to those before treatment (blue circle symbols, \(\ell_\delta > 440\,\mathrm{nm}\)) indicates enhanced thermal conductivity following the mid-T heat treatment of niobium cavities. This observation is consistent with our recent study on the thermal conductivity of mid-T heat-treated niobium \mbox{samples}~\cite{Wenskat2025}.}
\end{figure}

 One of the key physical parameters in frequency-shift measurements is the critical temperature. However, due to spatial temperature gradients along the cavities during the measurements, the associated uncertainties have so far been too large to allow a reliable determination of this parameter. To illustrate the impact of spatial temperature gradients \(\nabla T\) on the determination of \(T_\mathrm{c}\), Fig.~\ref{Fig_09} presents the critical temperatures extracted from both \(f(T)\) and \(Q_\mathrm{L}(T)\) for several baseline cavities with large mean free paths within the effective \mbox{normal-conducting} skin depth \(\ell_\delta\) as well as for cavities subjected to mid-T heat treatment with reduced \(\ell_\delta\) values (see Appendix for the calculation of \(\ell_\delta\)). Cavities treated either by electropolishing (EP) or by annealing at $800^{\circ}\mathrm{C}$ for 3~hours are referred to as baseline cavities. For the \mbox{mid-T} \mbox{heat-treated} cavities, the dip-onset temperature \(T_\mathrm{onset}\) in \(f(T)\) curve is included for comparison (see Fig.~\ref{Fig_08}). A precise definition of the critical temperature for curves exhibiting the dip phenomenon requires further investigation.

\par The presence of a spatial temperature gradient along the cavity shifts the apparent \(T_\mathrm{c}\) to higher values. As shown in Fig.~\ref{Fig_09}, the \(T_\mathrm{c}\) of baseline cavities (\(\ell_\delta > 440\,\mathrm{nm}\)) shifts more strongly with increasing gradient than that of mid-T heat-treated cavities (\(\ell_\delta < 75\,\mathrm{nm}\)), indicating that the temperature-gradient effect is more pronounced in the baseline cavities. A previous study by our group~\cite{Wenskat2025}, focusing on the thermal conductivity of niobium before and after mid-T heat treatment, demonstrated that this treatment enhances the thermal conductivity of niobium samples. Our observations in Fig.~\ref{Fig_09} are consistent with the findings reported in~\cite{Wenskat2025}: the reduced effect of the spatial temperature gradient on the shift of \(T_\mathrm{c}\) toward higher values, observed in mid-T heat-treated cavities compared to baseline cavities, is assumed to be attributable to their higher thermal conductivity.

\par Based on previous studies, a suppression of the \mbox{intrinsic} critical temperature is expected as a result of interstitial oxygen introduced into the niobium bulk~\cite{Desorbo1963}, under conditions where the spatial temperature gradient is negligible. Indeed, Fig.~\ref{Fig_08} illustrates this suppression of \(T_\mathrm{c}\) in the mid-T heat-treated cavity compared with its baseline state, as observed in both \(f(T)\) and \(Q_\mathrm{L}(T)\). In Fig.~\ref{Fig_09}, although the data exhibit relatively large uncertainties, the trend toward lower \(T_\mathrm{c}\) and \(T_\mathrm{onset}\) values in the mid-T heat-treated cavities relative to the baseline cavities qualitatively indicates the expected suppression of \(T_\mathrm{c}\).

\begin{figure}
\includegraphics[scale=0.345]{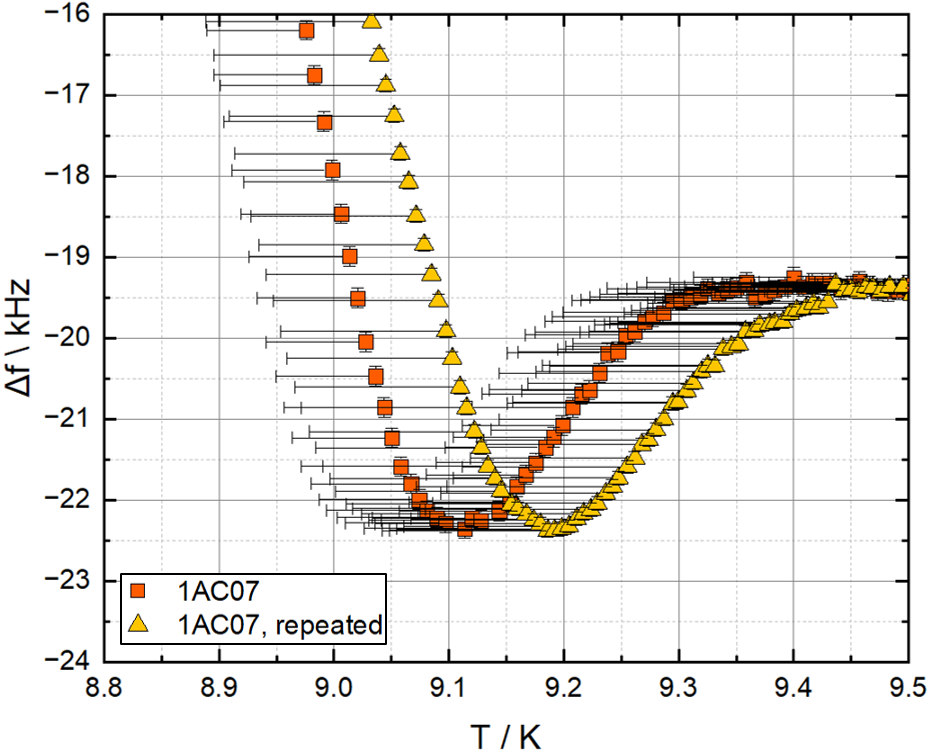}
\caption{\label{Fig_10}
Magnified temperature dependence of the \mbox{resonance-frequency} shift in the region of dip emergence around the critical temperature for a mid-T heat-treated \mbox{cavity} tested twice under different spatial temperature gradients. The effects of the spatial temperature gradient on the \mbox{cavity-temperature} readings and the corresponding uncertainties in the frequency shift are shown as error bars. These results confirm that, although the analysis of absolute temperature values is limited by high uncertainty, the influence of the spatial temperature gradient on the characterization of the dip---as a narrow relative feature---is negligible.}
\end{figure}

\par Regardless of the temperature-gradient effect, higher \(T_\mathrm{c}\) values are obtained from \(Q_\mathrm{L}(T)\) than from \(f(T)\) for each baseline-cavity test shown in Fig.~\ref{Fig_09}. In Fig.~\ref{Fig_08}, corresponding to the measurement with the most reliable dataset in our experiments---featuring the lowest \(\nabla T\) of approximately \(2\,\mathrm{K/m}\), as shown in Fig.~\ref{Fig_09}---the baseline cavity yields a \(T_\mathrm{c}\) value of about \(9.29\,\mathrm{K}\) derived from \(f(T)\), which agrees well with the literature value of \(9.27\,\mathrm{K}\) for niobium cavities. However, the \(T_\mathrm{c}\) value obtained from \(Q_\mathrm{L}(T)\) appears slightly higher (\(\sim 9.35\,\mathrm{K}\)). This observation suggests that the \(T_\mathrm{c}\) determined from \(f(T)\) may serve as a more precise and reliable indicator than that obtained from \(Q_\mathrm{L}(T)\). In contrast, for each mid-T heat-treated cavity test, the \(T_\mathrm{c}\) value derived from the \(Q_\mathrm{L}(T)\) curve is lower than the \(T_\mathrm{onset}\) value obtained from the \(f(T)\) curve. This behavior may arise from one or both of the following reasons: first, the effect of thermal conductivity differs in mid-T heat-treated cavities compared with baseline cavities and may influence \(Q_\mathrm{L}(T)\) and \(f(T)\) differently; second, \(T_\mathrm{onset}\) may not represent the true critical temperature in \(f(T)\) curves exhibiting a dip. An alternative candidate for the definition of the critical temperature in the \(f(T)\) curve of mid-T \mbox{heat-treated} cavities could correspond to the temperature at which the frequency dip reaches its minimum. However, further investigations under minimized \(\nabla T\) are essential to clarify the critical temperature.

\par Interestingly, although large spatial temperature gradients increase the overall uncertainty in the \mbox{cavity-temperature} readings and complicate the determination of \(T_\mathrm{c}\), their effect within local, relative temperature ranges---such as the narrow region encompassing the dip---is largely unidirectional and nearly constant across that range. As a result, the influence of the temperature gradient on the intrinsic shape and relative characteristics of the dip, such as its frequency magnitude and temperature width, is minimal and effectively \mbox{cancels} out. This observation has been confirmed through repeated measurements, and one exemplary case is shown in Fig.~\ref{Fig_10}. As illustrated in the figure, even under conditions of significant and different temperature uncertainties for each test, the entire dip shifts while its relative characteristics remain largely unaffected. Hence, we assume that the effect of the spatial temperature gradient on the relative characteristics of the dip feature is negligible. Therefore, the analysis of the dip characteristics remains physically meaningful and reliable for all previously measured data in our experiments.
\section{Conclusion}
\par Implementing frequency-shift measurements for \mbox{single-cell} cavities under cryogenic, \mbox{non-adiabatic} \mbox{warm-up} conditions in large cryostats designed for \mbox{nine-cell} cavities presents significant technical challenges, leading to comparatively high uncertainties in both the resonance frequency and its temperature dependence. In this work, these limitations are addressed through a consistent measurement and analysis framework combined with a systematic uncertainty evaluation, enabling the determination of average electron mean free paths in both the superconducting and \mbox{normal-conducting} regimes. In addition, frequency-dip features emerging near the critical temperature in the resonance-frequency-shift behavior of mid-T heat-treated cavities are reliably identified and characterized. These results constitute an important step toward a deeper understanding of the underlying physical mechanisms and provide a solid basis for further analytical developments within ongoing research.

\par More recently, reducing the effect of mechanical stress on the cavity resonance frequency during measurement, combined with a practical method for minimizing spatial temperature gradients along the cavity, has led to major improvements in both the reproducibility and precision of the frequency-shift measurements. These advances, which are still under active development, provide a robust and accurate basis for temperature-dependent studies of the complex impedance above 7\,K in SRF cavities. With the established methodology, our frequency-shift commissioning now fulfills the systematic requirements for comprehensive fundamental electromagnetic studies based on resonance-frequency-shift measurements.

\begin{acknowledgments}
The authors would like to thank S. Barbanotti, K. Jensch, and the clean-room and AMTF teams at DESY for preparing the cavities and performing the vertical tests, and to express their gratitude to the University of Hamburg SRF team. This work was partially funded by the Helmholtz Association within the topic Accelerator Research and Development (ARD) of the Matter and Technology (MT) Program and was supported by the BMBF project under research grants 05K19GUB, 05H2021, and 05H2024.
\end{acknowledgments}
\section*{Appendix}
\par At elevated temperatures, where \(Q_\mathrm{L}\) further decreases, the relative contribution of input antenna coupling losses becomes increasingly insignificant compared to the intrinsic RF losses of the cavity. Above \(T_\mathrm{c}\), the measured \(Q_\mathrm{L}\) typically falls within the range of \(10^4\)\textendash\(10^5\), indicating that the input-coupling term in Eq.~\ref{Eq_05} is negligible in this regime, as \(Q_{\mathrm{ext,pk}}\) and \(Q_{\mathrm{ext,ant}}\) are on the order of \(10^{11}\) and \(10^{8}\), respectively. Consequently, \(Q_0\) and \(Q_\mathrm{L}\) can be considered effectively equivalent (see Fig.~\ref{Fig_08}), and the \(Q_\mathrm{L}\) values above \(T_\mathrm{c}\) can be reliably used for further cavity characterization. Given that the critical temperature of niobium is \(9.27\,\mathrm{K}\), the surface resistance in the normal-conducting region is determined using \(Q_0\) at 10\,K, according to Eq.~\ref{Eq_01}, defined as
\begin{equation}
\label{Eq_07}
R_\mathrm{s}(10\,\mathrm{K}) = \frac{G}{Q_0(10\,\mathrm{K})}.
\end{equation}

\par Additionally, the total frequency shift observed in the \(f(T)\) curve represents another reliable quantity that can be used for further analysis when its associated uncertainty is taken into account. The total effective superconducting penetration depth, \(\Delta \lambda_\mathrm{max}\), is determined from the measured \(\Delta f_\mathrm{max}\) (see Fig.~\ref{Fig_08}) using Eq.~\ref{Eq_03}, expressed as
\begin{equation}
\label{Eq_08}
\Delta \lambda_\mathrm{max} = \frac{-G \, \Delta f_\mathrm{max}}{\pi \mu_0 f^2}.
\end{equation}
The resulting values of \(R_\mathrm{s}(10\,\mathrm{K})\) and \(\Delta \lambda_\mathrm{max}\) serve as key parameters for investigating the electron mean free path \(\ell\) and for quantitatively assessing the impurity concentration in the near-surface region of the cavities.

\par The normal-conducting surface resistance of the cavities is analyzed within the framework of the normal and anomalous skin effects above the critical temperature, as described in Ref.~\cite{Padamsee1998}. The average mean free path in the effective normal-conducting skin depth (\(\ell_{\delta}\)) can be calculated directly in the dirty and clean limits, as given by
\begin{align}
\label{Eq_09}
\ell_{\delta}\hspace{-0.2em}&=\hspace{-0.2em} 
    \begin{cases}
        \frac{\pi {f} \mu\, \rho\hspace{-0.06em}\ell}{R_\mathrm{s}^2}, & \scriptstyle\text{if }  R_\mathrm{s} \geq 1.9  R_\mathrm{s}(\infty),  \\
        {\scriptstyle1.042}\hspace{-0.15em}\left( \frac{\rho\hspace{-0.06em}\ell}{\pi f \, \mu} \right)^{1/3}\hspace{-0.2em} \left( \frac{R_\mathrm{s}}{R_\mathrm{s}(\infty)} - 1 \right)^{-1.209}, & \scriptstyle\text{if } R_\mathrm{s} < 1.9 R_\mathrm{s}(\infty),
    \end{cases}
\end{align}
where \(\mu\) and \(\rho\) represent the magnetic permeability and resistivity of the material, respectively. For niobium, \(\mu \approx \mu_0\). The resistivity is related to the electron mean free path through the \(\rho\ell\) product, a material constant that is independent of temperature. For niobium, the \(\rho\ell\) product is \(3.7 \times 10^{-16}~\Omega\cdot\mathrm{m}^2\)~\cite{Garwin1972}. In theory, the results of the anomalous skin effect are characterized by a dimensionless parameter defined as
\begin{equation}
\label{Eq_10}
 \alpha_\mathrm{s} = \frac{3}{2}\pi\mu f \frac{\ell_\delta^3}{\rho\hspace{-0.06em}\ell}.  
\end{equation}
Intermediate values of the normal surface resistance occur where \(\alpha_\mathrm{s} \geq 3\), which leads to the limit \mbox{\(R_\mathrm{s} \geq 1.9  R_\mathrm{s}(\infty)\)}, where \(R_\mathrm{s}(\infty)\) represents the surface resistance in the anomalous limit:
\begin{equation}
\label{Eq_11}
R_\mathrm{s}(\infty) = 1.29 \times 10^{-4} f^{2/3} \, (\rho\hspace{-0.06em}\ell)^{1/3}.
\end{equation}
This results in a resistance limit of \(1.1 \,\mathrm{m}\Omega\) for an extremely clean Nb 1.3\,GHz cavity, and thus the threshold \(R_\mathrm{s} \approx 2.1 \,\mathrm{m}\Omega\) marks the intermediate point at which the normal skin effect should apply for an Nb 1.3\,GHz cavity. Ultimately, \(\ell_{\delta}\) values for all measured cavities can be calculated using Eq.~\ref{Eq_09}.

\par To provide a parameter for comparing the surface quality of the cavities with reported values, the residual resistance ratio in the effective normal-conducting skin-depth range, \(RRR_\delta\), can be defined as the ratio of the surface resistivity at room temperature, \(\rho(300\,\mathrm{K})\), to the surface resistivity at 10\,K, \(\rho_\delta\), expressed as
\begin{equation}
\label{Eq_12}
 RRR_{\delta} = \frac{\rho(300\,\mathrm{K})} {\rho_{\delta}(10\,\mathrm{K})}, 
\end{equation}
where \(\rho_{\delta}\) is derived as \(\rho\hspace{-0.06em}\ell/\ell_\delta\). To evaluate the surface resistivity at room temperature, a subset of ten cavities was measured using the VNA and Eq.~\ref{Eq_04}, yielding an average resistivity value of \(1.37 \times 10^{-7} \, \Omega \, \mathrm{m}\). This value is consistent with literature values for niobium at room temperature~\cite{Padamsee1998}. The average \(RRR_\delta\) value of approximately 200 calculated for baseline cavities is lower than the typical bulk value of \(\sim 300\) for the Nb sheets used to fabricate the cavities. The \(RRR_\delta\) derived from the surface resistivity relates specifically to the effective normal-conducting skin-depth properties of the cavities and does not reflect their bulk properties. Moreover, the skin depth \(\delta\) can be calculated using
\begin{equation}
\label{Eq_13}
 \delta = \frac{\rho_\delta }{R_\mathrm{s}(10\,\mathrm{K})}. 
\end{equation}

\par The dependence of \(\lambda_0\) in Eq.~\ref{Eq_07} on \(\ell\) is given by~\cite{Ip1948, Poole2007}:
\vspace{-5pt}
\begin{equation}
\label{Eq_14}
\lambda_0(\ell) = \lambda_\mathrm{L} \sqrt{1 + \frac{\pi \xi_0}{2\ell}},
\end{equation}
where \(\lambda_\mathrm{L}\) and \(\xi_0\) represent the London penetration depth at zero temperature and the coherence length of Cooper pairs in a pure material, respectively; for pure niobium, both are typically reported to be \(\sim 38\,\mathrm{nm}\).

\par The dependence of \(\lambda\) on \(T\) and the average mean free path in the superconducting region, \(\ell_\lambda\), is described using Eqs.~\ref{Eq_02} and~\ref{Eq_14}:
\begin{equation}
\label{Eq_15}
 \lambda(T, \ell_\lambda) = \lambda_\mathrm{L} \sqrt{1 + \frac{\pi \xi_0}{2\ell_{\lambda}}} \left[1 - \left(\frac{T}{T_\mathrm{c}}\right)^4 \right]^{-1/2}.
\end{equation}
By selecting a fixed temperature range, specifically from \(7\,\mathrm{K}\) to \(T_\mathrm{c}\), for the definition of \(\Delta \lambda_{\mathrm{max}}\) in this study, the temperature dependence in Eq.~\ref{Eq_15} is eliminated through a comparative analysis of different cavities, allowing us to focus solely on the mean-free-path dependence. Accordingly, we assume that any observed differences in the total effective superconducting penetration depth among mid-T heat-treated cavities arise from variations in \(\ell_\lambda\). This assumption helps isolate the role of \(\ell_\lambda\) in \(\Delta\lambda_{\mathrm{max}}\) differences and, based on Eq.~\ref{Eq_15}, allows us to consider the expression
\begin{equation}
\begin{aligned}
\label{Eq_16}
&\frac{\Delta\lambda_\mathrm{max}}{\sqrt{1 + \frac{\pi \xi_0}{2\ell_{\lambda}}}} 
= \frac{\lambda(\sim T_\mathrm{c}) - \lambda(7\,\mathrm{K})}{\sqrt{1 + \frac{\pi \xi_0}{2\ell_{\lambda}}}} = \text{const.} \\
&= \lambda_\mathrm{L}\!\left[1 - \left(\frac{\sim T_\mathrm{c}}{T_\mathrm{c}}\right)^{\!4}\right]^{\!-1/2}
 - \lambda_\mathrm{L}\!\left[1 - \left(\frac{7\,\mathrm{K}}{T_\mathrm{c}}\right)^{\!4}\right]^{\!-1/2},
\end{aligned}
\end{equation}
to be constant across different treatments.

\par At this stage, we have chosen to estimate \(\ell_{\lambda}\) using the constant expression in Eq.~\ref{Eq_16} for \(\Delta\lambda_{\mathrm{max}}\). This approach is adopted because fitting Eq.~\ref{Eq_15} directly to the \(\Delta f(T)\) curve does not yield an accurate estimate of \(\ell_{\lambda}\) for \mbox{mid-T} heat-treated cavities, even when \mbox{high-precision} curves are available. The dip phenomenon observed below \(T_\mathrm{c}\) modifies the behavior of the \(\Delta f(T)\) curve in \mbox{mid-T} \mbox{heat-treated} cavities relative to baseline cavities. In Fig.~\ref{Fig_08}(a), the dip behavior is illustrated through the resonance-frequency shift after a mid-T heat treatment. The curve behavior, including the emergence of the dip and its influence below the dip temperature range compared to before mid-T heat treatment, can not be explained using traditional frameworks such as the \mbox{Gorter--Casimir} model~\cite{Gorter1934}. Employing a novel theoretical framework is therefore essential for studying and estimating the superconducting parameters when fitting \(\Delta f(T)\) data exhibiting a dip, particularly for treatments that introduce interstitial oxygen or nitrogen into the Nb structure. Such work is currently under investigation by theorists~\cite{Ueki2023, Lebedeva2024}. The use of the constant expression in Eq.~\ref{Eq_16} follows the assumption that the mechanism causing the frequency-dip phenomenon around \(T_\mathrm{c}\) may lead to deviations in the \(f(T)\) curve near and below \(T_\mathrm{c}\), but does not affect the total magnitude of the frequency shift; therefore, \(\Delta\lambda_\mathrm{max}\) remains dependent solely on \(\ell_{\lambda}\).

\par To calculate the constant magnitude of the expression in Eq.~\ref{Eq_16}, an experimental reference is required. The values of \(\Delta\lambda_\mathrm{max}\) can be extracted for each \(f(T)\) curve. However, as no reference values for \(\ell_{\lambda}\) were available, a substitute had to be found. Since \(\ell_{\delta}\) can be derived from the measured \(R_\mathrm{s}\) values, and it is known that baseline cavities exhibit a nearly constant impurity concentration with negligible variation across both the effective superconducting penetration depth and the normal-conducting skin depth, the average of the calculated \(\ell_{\delta}\) values from the normal surface resistance of eight baseline cavities was used. Thus, the expression
\begin{equation}
\label{Eq_17}
\frac{\Delta\lambda_{\mathrm{max}}} {\sqrt{1 + \frac{\pi \xi_0}{2\ell_{\lambda}}}} \cong 219 \pm 16 ~\mathrm{nm}
\end{equation}
was obtained on average, using the \(\ell_{\delta}\) of the baseline cavities instead of their \(\ell_{\lambda}\) and using their \(\Delta\lambda_\mathrm{max}\) in Eq.~\ref{Eq_17}. For \(\xi_0\), the literature value of \(39\,\mathrm{nm}\) for niobium was applied. Assuming that the expression in Eq.~\ref{Eq_17} remains constant because of the fixed temperature range defined in Eq.~\ref{Eq_16}, the value \(219 \pm 16 \,\mathrm{nm}\) can be used for all cavities, whether heat-treated or not, to determine \(\ell_{\lambda}\) from their corresponding \(\Delta\lambda_{\mathrm{max}}\), accounting for the associated errors.

\par Since the resistivity prior to mid-T heat treatment is negligibly small compared to that after the treatment, together with the underlying assumption that the resistivity after mid-T heat treatment arises from interstitial oxygen in bulk niobium, the average oxygen concentration in mid-T heat-treated cavities within the effective normal-conducting skin depth, \(C_{\mathrm{O},\delta}\), can be defined as
\begin{equation}
\label{Eq_18}
\hspace{5em} C_{\mathrm{O},\delta} = \frac{\rho_{\delta}}{4.5\times 10^{-8}~/~\Omega\cdot\mathrm{m}\cdot\mathrm{at.\%}^{-1}}.
\end{equation}
Moreover, the average oxygen concentration in the effective superconducting penetration-depth range, \(C_{\mathrm{O},\lambda}\), can be calculated by using \(\rho_{\lambda}\), which can be derived as \(\rho\hspace{-0.06em}\ell/ \ell_\lambda\), and applying it to the oxygen-concentration--resistivity relationship, defined as
\begin{equation}
\label{Eq_19}
C_{\mathrm{O},\lambda} = \frac{\rho_{\lambda}}{4.5\times 10^{-8}~/~\Omega\cdot\mathrm{m}\cdot\mathrm{at.\%}^{-1}}. 
\end{equation}
Therefore, it becomes possible to investigate the average concentration of interstitial oxygen within both the effective total superconducting penetration depth and the effective normal-conducting skin depth.
\bibliography{apssamp}
\end{document}